\documentclass[twocolumn]{webofc}
\usepackage[varg]{txfonts}   
\newcommand{\newc}{\newcommand}
\newc{\beq}    {\begin{equation}}
\newc{\eeq}    {\end{equation}}

\newc{\beqa}    {\begin{eqnarray}}
\newc{\eeqa}    {\end{eqnarray}}
\newc{\bs}    {\section}
\newc{\no}    {\\ \nonumber}

\def\apj{{\em Astrophys. J.  }}
\def\apjl{{\em Astrophys. J. Lett. }}

\def\nat{Nature }

\def\pra{{ Phys. Rev.} {\bf A} }
\def\prd{{ Phys. Rev.} {\bf D} }

\def\mnras{{ Mon. Not. Roy. Astron. Soc.  }}

\newc{\st}    {\stackrel}

\begin{document}
\title{Brief History of Ultra-light Scalar  Dark Matter Models}
%
%

\author{\firstname{Jae-Weon } \lastname{Lee}\inst{1}\fnsep\thanks{\email{scikid@jwu.ac.kr}}
}

\institute{Department of renewable energy, Jungwon university,
            85 Munmu-ro, Goesan-eup, Goesan-gun, Chungcheongbuk-do,
              367-805, Korea }

\abstract{This is a  review on the brief history of the scalar field dark matter model
also known as fuzzy dark matter, BEC dark matter, wave dark matter, or ultra-light axion.
 In this model ultra-light scalar dark matter particles with mass $m = O(10^{-22})eV$
condense in a single Bose-Einstein condensate  state and behave collectively like a classical wave.
Galactic dark matter halos can be described
as a self-gravitating coherent scalar field configuration called boson stars.
 At the scale larger than galaxies the
dark matter acts like cold dark matter, while below the scale quantum pressure from the uncertainty principle suppresses the
 smaller structure formation so that it can  resolve the small scale crisis of the conventional
 cold dark matter model.
}
\maketitle

Despite long efforts dark matter (DM) remains a great mystery in physics and astronomy~\cite{Silk:2016srn}.
While numerical results of the  cold dark matter (CDM)  model  are
remarkably successful in explaining the large scale structure of the universe, it
encounters some problems at the scale of  galactic or sub-galactic
structures. For example, the numerical simulations  usually predict cusped central halo density and
too many sub-halos and small galaxies, which seem to be in contradiction with observations~\cite{Salucci:2002nc,navarro-1996-462,deblok-2002,crisis}.

Recently, there is a growing  interest in the idea that
DM particles are  ultra-light scalars  in Bose-Einstein condensate (BEC).
(For a review, see Refs. \citealp{2009JKPS...54.2622L,2014ASSP...38..107S,2014MPLA...2930002R,2014PhRvD..89h4040H,2011PhRvD..84d3531C,2014IJMPA..2950074H,Marsh:2015xka,Hui:2016ltb})
In this model, due to the tiny DM particle mass $m = O(10^{-22})eV$, the DM particle number density is very high and hence
the inter-particle distance is much smaller than the de Broglie wave length of the DM particles. Therefore, the particles are in BEC and
move collectively as a wave rather than incoherent particles.
 At the scale larger than galaxies the
DM perturbation behaves like that of CDM, while below the scale quantum pressure from the uncertainty principle suppresses the
 small structure formation, which makes it a viable alternative to CDM.
This property helps us to resolve the small scale problems of the CDM model such as the missing satellite problem or
the cusp/core problem~\cite{2009NJPh...11j5029P}.

Before 2000 there were only a few groups of scientists  working on this topic, without much communication
 even between them.
Being  unaware of precedent works,  many researchers in this field
 independently proposed  similar ideas with various names such as
   BEC DM, scalar field DM (SFDM),  fuzzy DM, ultra-light axion (ULA), ultra-light axion like particle (ALP), wave DM,
$\psi$ DM, repulsive DM  or (super)fluid DM among many, although
the basic physics of these models is quite similar. (Henceforth, I will use the term 'SFDM' for the model.)
This unfortunate  situation has brought some confusions
among researchers  in this field.
Therefore, at this point, it is  desirable to summarize  what have  already attempted in this exciting field. \footnote{The list
in the references of this paper is by no means  exhaustive.}

The hypothesis that galactic DMs are   ultra-light scalar particles  in BEC has a long history.
In Ref. ~\cite{1983PhLB..122..221B} Baldeschi  et al.  considered
    the galactic halo model of
self-gravitating bosons with mass $m\simeq 10^{-24} eV$, which was obtained by comparing the de Broglie wave length of DM
to the typical galaxy size.
In ~\cite{1989PhRvA..39.4207M} Membrado  et al.  calculated
 the rotation curve
of self-gravitating boson halos  using the ground state of the bosons, which later turned out to be adequate for dwarf galaxies
 \cite{Matos:2003pe}.
In  \cite{Sin:1992bg}  Sin  suggested that the halos
are   like  giant atoms made of
ultra light BEC DM particles such as
pseudo Nambu-Goldstone boson (PNGB). He tried to explain  the observed  flat rotation curves (RCs)  using the excited states of the BEC DM
 of the
 Non-linear Schr$\ddot{o}$dinger equation  and
  emphasized the ripple structures of the RCs as a smoking gun of the theory. The particle mass $m\simeq 3 \times 10^{-23} eV$ was first obtained by fitting
  the observed RC of  galaxy NGC2998. (See also \citep{sin2,sin3})
 Following his work, Lee and Koh ~\cite{myhalo,kps} suggested that   DM halos are   giant boson stars (BSs) which can be
  described by the relativistic scalar field theory with gravity, i.e., Einstein Klein-Gordon equation (EKG).
In this  model DM particles are represented by
   scalar field $\phi$
with a typical action
\beq
\label{action}
 S=\int \sqrt{-g} d^4x[\frac{-R}{16\pi G}
-\frac{g^{\mu\nu}} {2} \phi^*_{;\mu}\phi_{;\nu}
 -U(\phi)],
\eeq
where the   potential
$U(\phi)=\frac{m^2}{2}|\phi|^2+\frac{\lambda}{4}|\phi|^4$ was considered as an  example. This model can be reduced to the BEC model  in the Newtonian limit.
The field is often assumed to be complex but can be real and have more general potentials.
From the maximum stable central density
and the maximum halo mass ($M_h\simeq 10^{12} M_\odot$)
  the mass bound $10^{-28} \leq  m \leq 10^{-22} eV$ was obtained for $\lambda=0$ case (now often called fuzzy DM)
from the BS theory.
It was emphasized that the repulsive self-interaction among DM particles, if there is,
provides an additional pressure against the gravitational collapse and drastically changes the length scale
to $O(\sqrt{\lambda} m_P/m^2)$
even for a tiny $\lambda$, where $m_P$ is the planck mass. An approximate analytic solution of KGE, $|\phi|^2\sim sin(kr)/kr$, was obtained
for $\Lambda\equiv {\lambda} m^2_P/m^2 \gg 1$ (Thomas-Fermi limt) \cite{myhalo,Boehmer:2007um}.
If $\lambda<0$ (as in the Taylor-expanded potential  for ULA) the halo can be unstable ~\cite{myhalo} as  argued
 in \cite{Guth:2014hsa}.

On the other hand Press et al.~\cite{1990PhRvL..64.1084P}, consider a `soft boson' with somewhat longer
Compton wavelength $\lambda_c=30 kpc$  than that for the typical SFDM.
Widrow and Kaiser~\cite{1993ApJ...416L..71W} proposed a numerical technique for the evolution of collisionless matter
using the  Schr\"{o}dinger-Poisson equations (SPE).

Similar ideas were independently investigated  by many authors  ~\cite {1993ApJ...416L..71W,Schunck:1998nq, PhysRevLett.84.3037,PhysRevD.64.123528,repulsive,fuzzy,
 corePeebles,Nontopological,Mielke2009174,PhysRevD.62.103517,Alcubierre:2001ea,2012PhRvD..86h3535P,2009PhRvL.103k1301S,
 Fuchs:2004xe,Matos:2001ps,0264-9381-18-17-101,PhysRevD.63.125016,Julien,Boehmer:2007um, 2012arXiv1212.5745B,Matos:2001ps,Eby:2015hsq}.
For example, a model with $m=0$ was suggested in ~\cite{Schunck:1998nq}, which has a stability issue. The SFDM without self-interaction (i.e., Fuzzy DM) with $m\simeq 10^{-22}eV$ was shown to be able to solve  the small scale issues~\cite{fuzzy}.
Guzman and Matos  et al. \cite{Matos:1998vk,Guzman:1998vg} extensively studied various astrophysical aspects of the SFDM.
Other attempts are based, for example, on a non-minimal coupling ~\cite{PhysRevLett.84.3037},
quintessence~\cite{PhysRevD.64.123528,PhysRevD.65.083514},
Repulsive DM ~\cite{repulsive}, fluid DM ~\cite{corePeebles}, nontopological soliton  ~\cite{Nontopological},
$cosh$ potential~\cite{PhysRevD.62.103517,Alcubierre:2001ea}, and ULA
~\cite{Mielke2009174,2009PhRvL.103k1301S,2012PhRvD..86h3535P}
among many ~\cite{Fuchs:2004xe,Matos:2001ps,0264-9381-18-17-101,PhysRevD.63.125016,Julien,Sahni:1999qe,2012arXiv1212.5745B}.
The BEC nature of this DM was extensively studied in ~\cite{Boehmer:2007um,2014PhRvD..89h4040H,2012A&A...537A.127C}
by Harko,  Chavanis and their colleagues.
More recently cosmology of the ULA is thoroughly studied in ~\cite{Marsh:2010wq,Hlozek:2014lca} by Marsh et al.
The ULA has usually a cosine potential, which is  often approximated to be a quadratic potential.
In that case the calculation results are almost same to those of the fuzzy DM.
Other works are described in the followings.


In the SFDM model the macroscopic wave function of the halo DM is often described by the following  SPE,
\beqa
i\hbar \partial_{{t}} {\psi} &=&-\frac{\hbar^2}{2m} \nabla^2 {\psi} +m{V} {\psi}, \no
\nabla^2 {V} &=&{4\pi G} \rho_d,
\eeqa
which can be derived from the EKG for weak gravity ~\cite{PhysRevD.35.3640} using $\psi=\sqrt{m}\phi$ or from  the many-body formalism.
Here, the DM density is $\rho_d=m|\psi|^2=m^2|\phi|^2$ and $V$ is the self-gravitation potential.
Once we get $\psi$ from the SPE we can predict the astrophysical observables. For example, for a spherical halo the  rotation velocity
at radius $r$ is
$
 v_{rot}(r)=\sqrt{\frac{GM(r)}{r}},
$
where $M(r)=4\pi \int^r_0 r'^2 \rho_d(r') d{r}'$ is the mass within $r$.

To study
the cosmological structure formation in the fluid approach, it is useful to reduce the Schr\"{o}dinger equation  to
the Madelung euations ~\cite{2011PhRvD..84d3531C,2014ASSP...38..107S} using
\beq
\psi(r,t)=\sqrt{\rho(r,t)}e^{iS(r,t)}.
\label{madelung}
\eeq
Substituting Eq. (\ref{madelung}) into the Schr\"{o}dinger equation gives
a continuity equation
\beq
\frac{\partial \rho}{\partial t} + \nabla \cdot (\rho \textbf{v})=0
\label{continuity}
\eeq
and an Euler-like equation
\beq
\frac{\partial \textbf{v}}{\partial t} + (\textbf{v}\cdot \nabla)\textbf{v} +\nabla V
+\frac{\nabla p}{\rho} -\frac{\nabla Q}{m} =0
\label{euler}
\eeq
with  a quantum potential $Q\equiv\frac{\hbar^2}{2m}\frac{\Delta \sqrt{\rho}}{\sqrt{\rho}}$,
a fluid velocity $\textbf{v}\equiv \nabla S/2m$, and  the pressure  $p$ from a self-interaction (if any).
The quantum pressure term
${\nabla Q}/{m}$ is absent in the CDM models.
 We have ignored  the cosmic expansion for simplicity.
Perturbing the above equations
 around $\rho=\bar\rho$, $\textbf{v}=0$, and $V=0$
 gives an equation for density
 perturbation $\delta\rho\equiv \rho-\bar{\rho}$,
 \beq
  \frac{\partial^2 \delta\rho}{\partial t^2}+\frac{\hbar^2}{4m^2}\nabla^2 (\nabla^2 \delta \rho)
  -c^2_s \nabla^2 \delta\rho - 4\pi G \bar{\rho}\delta\rho=0,
 \eeq
 where $c_s$ is the classical sound velocity from $p$, and $\bar{\rho}$ is the average background matter density (See, for example, Refs. \citealp{2012A&A...537A.127C, Suarez:2011yf}
 for details.).

 The Fourier transformed equation of the
  density contrast $\delta\equiv\delta \rho/\bar{\rho}=\sum_k \delta_k e^{ik\cdot r}$ with a wave vector $k$ is
  \beq
  \frac{d^2 \delta_k}{d t^2} +  \left[(c^2_q+c^2_s)k^2-4\pi G \bar{\rho} \right]\delta_k=0,
 \eeq
 where $c_q=\hbar k/2m$ is a quantum velocity.
For a small $k$ (at a large scale) the $c_s$ dependent term dominates and  SFDM behaves like the CDM (with pressure)
 while for a large $k$ (at a small scale) the $c_q$ dependent term dominates and quantum pressure prevents the structure formation ~\cite{fuzzy,Alcubierre:2002et,PhysRevD.63.063506,Harko:2011jy}.
 This interesting behavior of the SFDM is confirmed by the precise numerical studies
 \cite{2014NatPh..10..496S} and makes the SFDM an ideal alternative for the CDM.
 This property  alleviates the small scale problems of the CDM model ~\cite{corePeebles,PhysRevD.62.103517,0264-9381-17-13-101,PhysRevD.63.063506}.
Equating $c^2_q k^2$ with $4\pi G \bar{\rho}$ gives the quantum Jeans length scale~\cite{1985MNRAS.215..575K,Grasso:1990zg,fuzzy} at the redshift $z$,
\beq
\label{lambdaQ}
\lambda_Q(z)= \frac{2\pi}{k}=\left(\frac{\pi^3 \hbar^2 }{m^2G\bar\rho(z)}\right)^{1/4}
\eeq%
which leads to the minimum length scale and the mass scale of galactic halos formed at $z$
~\cite{Lee:2008ux,Lee:2015cos,Lee:2008jp}.


Dwarf galaxies are the smallest dark-matter-dominated  objects and, hence,
 are ideal objects for understanding the nature of dark matter.
  The Jeans length  above provides a natural cut-off in the mass power spectrum preventing the overproduction of subhalos
 and the dwarf satellite galaxies   ~\cite{corePeebles,PhysRevD.62.103517,0264-9381-17-13-101,PhysRevD.63.063506,fuzzy,2011MNRAS.416...87S}.
The SFDM could solve the cusp problem  ~\cite{fuzzy,Matos:2003pe,corePeebles,Riotto:2000kh,Su:2010bj,corePeebles,Riotto:2000kh,Matos:2007zza}
because of the boundary condition for the SPE at the halo center.
There are many works explaining  the rotation curves of dwarf \cite{Matos:2003pe,PhysRevD.68.023511,Boehmer:2007um}, and large galaxies  \citep{Lesgourgues2002791,Robles:2012uy,Schive:2014hza,2011MNRAS.413.3095H,Julien,2012MNRAS.422..282R,2017arXiv170100912B,PhysRevD.64.123528,0264-9381-17-1-102,
Mbelek:2004ff,PhysRevD.69.127502} in this model.
By fitting RCs one can obtain the scalar field mass $m \sim 10^{-22}eV$, and
interestingly the SFDM with this mass range can resolve many of the small scale problems of CDM models.
One exception is the Lyman alpha tension~\cite{Irsic:2017yje,Armengaud:2017nkf} which is under debate~\cite{Zhang:2017chj}.
This model can be used to explain the puzzling
 minimum mass scale~\cite{Lee:2008jp,Lee:2015cos} and the minimum length scale of galaxies~\cite{Mateo:1998wg,gilmore-2008} regardless of their luminosity  ~\cite{Strigari:2008ib} and
the  observed size  evolution ~\cite{Lee:2008ux} of the most massive galaxies ~\cite{Daddi:2005ym,2009Natur.460..717V,Trujillo21112007}.


The rotating ellipsoidal gravitational potentials of the SFDM halos was shown to
   induce spiral arms and shell structures in the visible matter of  galaxies~\cite{2012arXiv1212.5745B}.
   The effect of embedding a supermassive black hole in a SFDM halo was considered in \cite{UrenaLopez:2002du}.
It was also argued that this model can explain the M-sigma relation of supermassive black holes ~\cite{Lee:2015yws}.


 It is  suggested that the SFDM  can explain the  contradictory behaviors of DM in
  collisions of galaxy clusters~\cite{Lee:2008mq,CastanedaValle:2013ava,Paredes:2015wga,Schwabe:2016rze,Guzman:2016peo}.
  For a fast collision as observed in the Bullet cluster  two dark matter halos pass each other in a soliton-like way,
   while for a slow collision as observed in the Abell 520 the halos merge  due to the wave nature of the SFDM.
 This idea might explain the origin of the dark galaxy and the galaxy without dark matter ~\cite{Lee:2008mq} .


The SFDM model reproduces the evolution of the cosmological densities \citep{2009MNRAS.393.1359M} and
  the spectrum of the cosmic microwave background \citep{2010ApJ...721.1509R,2012PhRvD..86h3535P} if $m> 10^{-24} eV$.


The typical QCD axion has some difficulties to be a SFDM. It has $cos(\phi)$ potential which is
effectively attractive to $O(\phi^4)$, and  its mass $m_a\simeq 10^{-5}eV$ is too heavy, and it is a real field giving oscillatons instead of
boson stars ~\cite{myhalo}. On the other hand ultra-light axion (ULA) with mass $m_a\simeq 10^{-22}eV$ recently
attracts much interest, but if we consider only the quadratic potential for ULA,
ULA is almost identical to the typical SFDM or
Fuzzy DM. Therefore, we can  use the cosmological  constraints of ULA for SFDM and vice versa in many situations.
There is an  idea that a vector field in a modified gravity action could
be identified with a BEC ~\cite{moffat-2006}.
SFDM can be dark energy ~\cite{Arbey:2001qi,Matos:2009rw,Huang:2013spa,Gogberashvili:2017gru} and have quantum entanglement ~\cite{Lee:2015ema}.


 It is also suggested that the time dependent potential induced by oscillating SFDM
  influences the pulse arrival residual which can be observed by the Square Kilometre Array (SKA) experiment ~\cite{Khmelnitsky:2013lxt, Aoki:2016mtn} or by  laser interferometer gravitational wave detectors ~\cite{Aoki:2016kwl}.
  The oscillation of SFDM might resonate with binary pulsars  ~\cite{Blas:2016ddr}.

  In summary, SFDM with mass about $10^{-22}eV$ satisfies many cosmological constraints~\cite{Li:2013nal}
  and seems to be a viable alternative to CDM. Therefore, this model  deserves careful considerations.
  Especially, it is desirable to find a good particle physics model for SFDM~\cite{Kim:2015yna,Hui:2016ltb}.

\subsection*{acknowledgments}
This work was supported by the Jungwon University Research Grant (2016-040).
\bibliographystyle{woc}

\begin{thebibliography}{108}

\bibitem{Silk:2016srn}
J.~Silk, \emph{{Challenges in Cosmology from the Big Bang to Dark Energy, Dark
  Matter and Galaxy Formation}}, in \emph{{14th International Symposium on
  Nuclei in the Cosmos (NIC-XIV) Niigata, Japan, June 19-24, 2016}} (2016),
  \texttt{1611.09846}

\bibitem{Salucci:2002nc}
P.~Salucci, F.~Walter, A.~Borriello, Astron. Astrophys. \textbf{409}, 53
  (2003), \texttt{astro-ph/0206304}

\bibitem{navarro-1996-462}
J.F. Navarro, C.S. Frenk, S.D.M. White, \apj \textbf{462}, 563 (1996)

\bibitem{deblok-2002}
W.J.G. {de Blok}, A.~Bosma, S.S. McGaugh, astro-ph/0212102  (2002)

\bibitem{crisis}
A.~Tasitsiomi, International Journal of Modern Physics D \textbf{12}, 1157
  (2003)

\bibitem{2009JKPS...54.2622L}
J.W. {Lee}, Journal of Korean Physical Society \textbf{54}, 2622 (2009),
  \texttt{0801.1442}

\bibitem{2014ASSP...38..107S}
A.~{Su{\'a}rez}, V.H. {Robles}, T.~{Matos}, Astrophysics and Space Science
  Proceedings \textbf{38}, 107 (2014), \texttt{1302.0903}

\bibitem{2014MPLA...2930002R}
T.~{Rindler-Daller}, P.R. {Shapiro}, Modern Physics Letters A \textbf{29},
  1430002 (2014), \texttt{1312.1734}

\bibitem{2014PhRvD..89h4040H}
T.~{Harko}, \prd \textbf{89}, 084040 (2014), \texttt{1403.3358}

\bibitem{2011PhRvD..84d3531C}
P.H. {Chavanis}, \prd \textbf{84}, 043531 (2011), \texttt{1103.2050}

\bibitem{2014IJMPA..2950074H}
K.~{Huang}, C.~{Xiong}, X.~{Zhao}, International Journal of Modern Physics A
  \textbf{29}, 1450074 (2014), \texttt{1304.1595}

\bibitem{Marsh:2015xka}
D.J.E. Marsh, Phys. Rept. \textbf{643}, 1 (2016), \texttt{1510.07633}

\bibitem{Hui:2016ltb}
L.~Hui, J.P. Ostriker, S.~Tremaine, E.~Witten (2016), \texttt{1610.08297}

\bibitem{2009NJPh...11j5029P}
J.R. {Primack}, New Journal of Physics \textbf{11}, 105029 (2009),
  \texttt{0909.2247}

\bibitem{1983PhLB..122..221B}
M.R. {Baldeschi}, G.B. {Gelmini}, R.~{Ruffini}, Physics Letters B \textbf{122},
  221 (1983)

\bibitem{1989PhRvA..39.4207M}
M.~{Membrado}, A.F. {Pacheco}, J.~{Sa{\~n}udo}, \pra \textbf{39}, 4207 (1989)

\bibitem{Matos:2003pe}
T.~Matos, D.~Nunez, astro-ph/0303455  (2003), \texttt{astro-ph/0303455}

\bibitem{Sin:1992bg}
S.J. Sin, Phys. Rev. \textbf{D50}, 3650 (1994), \texttt{hep-ph/9205208}

\bibitem{sin2}
S.U. Ji, S.J. Sin, Phys. Rev. D \textbf{50}, 3655 (1994)

\bibitem{sin3}
C.~Lee, S.~Sin, J.Korean Phys.Soc. \textbf{28}, 16 (1995)

\bibitem{myhalo}
J.W. Lee, I.G. Koh, Phys. Rev. \textbf{D53}, 2236 (1996),
  \texttt{hep-ph/9507385}

\bibitem{kps}
J.W. Lee, I.G. Koh, {\it Galactic halo as a soliton star }, Abstracts, bulletin
  of the Korean Physical Society, 10 (2)  (1992)

\bibitem{Boehmer:2007um}
C.G. Boehmer, T.~Harko, JCAP \textbf{0706}, 025 (2007), \texttt{0705.4158}

\bibitem{Guth:2014hsa}
A.H. Guth, M.P. Hertzberg, C.~Prescod-Weinstein, Phys. Rev. \textbf{D92},
  103513 (2015), \texttt{1412.5930}

\bibitem{1990PhRvL..64.1084P}
W.H. {Press}, B.S. {Ryden}, D.N. {Spergel}, Physical Review Letters
  \textbf{64}, 1084 (1990)

\bibitem{1993ApJ...416L..71W}
L.M. {Widrow}, N.~{Kaiser}, \apjl \textbf{416}, L71 (1993)

\bibitem{Schunck:1998nq}
F.E. Schunck, astro-ph/9802258  (1998)

\bibitem{PhysRevLett.84.3037}
U.~Nucamendi, M.~Salgado, D.~Sudarsky, Phys. Rev. Lett. \textbf{84}, 3037
  (2000), \texttt{gr-qc/0002001}

\bibitem{PhysRevD.64.123528}
A.~Arbey, J.~Lesgourgues, P.~Salati, Phys. Rev. D \textbf{64}, 123528 (2001)

\bibitem{repulsive}
J.~Goodman, New Astronomy Reviews \textbf{5}, 103 (2000)

\bibitem{fuzzy}
W.~Hu, R.~Barkana, A.~Gruzinov, Phys. Rev. Lett. \textbf{85}, 1158 (2000)

\bibitem{corePeebles}
P.~Peebles, \apj \textbf{534}, L127 (2000)

\bibitem{Nontopological}
E.W. Mielke, F.E. Schunck, Phys. Rev. D \textbf{66}, 023503 (2002)

\bibitem{Mielke2009174}
E.W. Mielke, J.A.V. Perez, Physics Letters B \textbf{671}, 174  (2009)

\bibitem{PhysRevD.62.103517}
V.~Sahni, L.~Wang, Phys. Rev. D \textbf{62}, 103517 (2000)

\bibitem{Alcubierre:2001ea}
M.~Alcubierre, F.S. Guzman, T.~Matos, D.~Nunez, L.A. Urena-Lopez,
  P.~Wiederhold, Class. Quant. Grav. \textbf{19}, 5017 (2002),
  \texttt{gr-qc/0110102}

\bibitem{2012PhRvD..86h3535P}
C.G. {Park}, J.c. {Hwang}, H.~{Noh}, \prd \textbf{86}, 083535 (2012),
  \texttt{1207.3124}

\bibitem{2009PhRvL.103k1301S}
P.~{Sikivie}, Q.~{Yang}, Physical Review Letters \textbf{103}, 111301 (2009),
  \texttt{0901.1106}

\bibitem{Fuchs:2004xe}
B.~Fuchs, E.W. Mielke, Mon. Not. Roy. Astron. Soc. \textbf{350}, 707 (2004),
  \texttt{astro-ph/0401575}

\bibitem{Matos:2001ps}
T.~Matos, F.S. Guzman, L.A. Urena-Lopez, D.~Nunez, astro-ph/0102419  (2001)

\bibitem{0264-9381-18-17-101}
M.P. Silverman, R.L. Mallett, Classical and Quantum Gravity \textbf{18}, L103
  (2001)

\bibitem{PhysRevD.63.125016}
U.~Nucamendi, M.~Salgado, D.~Sudarsky, Phys. Rev. D \textbf{63}, 125016 (2001)

\bibitem{Julien}
A.A. Julien~Lesgourgues, P.~Salati, New Astronomy Reviews \textbf{46}, 791
  (2002)

\bibitem{2012arXiv1212.5745B}
H.L. {Bray}, ArXiv:1212.5745  (2012), \texttt{1212.5745}

\bibitem{Eby:2015hsq}
J.~Eby, C.~Kouvaris, N.G. Nielsen, L.C.R. Wijewardhana, arXiv:1511.04474
  (2015)

\bibitem{Matos:1998vk}
T.~Matos, F.S. Guzman, Class. Quant. Grav. \textbf{17}, L9 (2000),
  \texttt{gr-qc/9810028}

\bibitem{Guzman:1998vg}
F.S. Guzman, T.~Matos, H.~Villegas-Brena, Rev. Mex. Astron. Astrofis.
  \textbf{37}, 63 (2001), \texttt{astro-ph/9811143}

\bibitem{PhysRevD.65.083514}
A.~Arbey, J.~Lesgourgues, P.~Salati, Phys. Rev. D \textbf{65}, 083514 (2002)

\bibitem{Sahni:1999qe}
V.~Sahni, L.M. Wang, Phys. Rev. \textbf{D62}, 103517 (2000),
  \texttt{astro-ph/9910097}

\bibitem{2012A&A...537A.127C}
P.H. {Chavanis}, Astronomy and Astrophysics \textbf{537}, A127 (2012),
  \texttt{1103.2698}

\bibitem{Marsh:2010wq}
D.J.E. Marsh, P.G. Ferreira, Phys. Rev. \textbf{D82}, 103528 (2010),
  \texttt{1009.3501}

\bibitem{Hlozek:2014lca}
R.~Hlozek, D.~Grin, D.J.E. Marsh, P.G. Ferreira, Phys. Rev. \textbf{D91},
  103512 (2015), \texttt{1410.2896}

\bibitem{PhysRevD.35.3640}
R.~Friedberg, T.D. Lee, Y.~Pang, Phys. Rev. D \textbf{35}, 3640 (1987)

\bibitem{Suarez:2011yf}
A.~Suarez, T.~Matos, Mon. Not. Roy. Astron. Soc. \textbf{416}, 87 (2011),
  \texttt{1101.4039}

\bibitem{Alcubierre:2002et}
M.~Alcubierre, F.S. Guzman, T.~Matos, D.~Nunez, L.A. Urena-Lopez,
  P.~Wiederhold, \emph{{Scalar field dark matter and galaxy formation}}, in
  \emph{{Dark matter in astro- and particle physics. Proceedings, 4th
  Heidelberg International Conference, DARK 2002, Cape Town, South Africa,
  February 4-9, 2002}} (2002)

\bibitem{PhysRevD.63.063506}
T.~Matos, L.~Arturo Ure\~na L\'opez, Phys. Rev. D \textbf{63}, 063506 (2001)

\bibitem{Harko:2011jy}
T.~Harko, Mon. Not. Roy. Astron. Soc. \textbf{413}, 3095 (2011),
  \texttt{1101.3655}

\bibitem{2014NatPh..10..496S}
H.Y. {Schive}, T.~{Chiueh}, T.~{Broadhurst}, Nature Physics \textbf{10}, 496
  (2014), \texttt{1406.6586}

\bibitem{0264-9381-17-13-101}
T.~Matos, L.A. Urena-Lopez, Class. Quant. Grav. \textbf{17}, L75 (2000)

\bibitem{1985MNRAS.215..575K}
M.I. {Khlopov}, B.A. {Malomed}, I.B. {Zeldovich}, \mnras \textbf{215}, 575
  (1985)

\bibitem{Grasso:1990zg}
D.~Grasso, Phys.Rev. \textbf{D41}, 2998 (1990)

\bibitem{Lee:2008ux}
J.W. Lee, Phys. Lett. \textbf{B681}, 118 (2009), \texttt{0805.2877}

\bibitem{Lee:2015cos}
J.W. Lee, Phys. Lett. \textbf{B756}, 166 (2016), \texttt{1511.06611}

\bibitem{Lee:2008jp}
J.W. Lee, S.~Lim, JCAP \textbf{1001}, 007 (2010), \texttt{0812.1342}

\bibitem{2011MNRAS.416...87S}
A.~{Su{\'a}rez}, T.~{Matos}, \mnras \textbf{416}, 87 (2011), \texttt{1101.4039}

\bibitem{Riotto:2000kh}
A.~Riotto, I.~Tkachev, Phys. Lett. \textbf{B484}, 177 (2000),
  \texttt{astro-ph/0003388}

\bibitem{Su:2010bj}
K.Y. Su, P.~Chen, JCAP \textbf{1108}, 016 (2011), \texttt{1008.3717}

\bibitem{Matos:2007zza}
T.~Matos, L.A. Urena-Lopez, Gen. Rel. Grav. \textbf{39}, 1279 (2007)

\bibitem{PhysRevD.68.023511}
A.~Arbey, J.~Lesgourgues, P.~Salati, Phys. Rev. D \textbf{68}, 023511 (2003)

\bibitem{Lesgourgues2002791}
J.~Lesgourgues, A.~Arbey, P.~Salati, New Astronomy Reviews \textbf{46}, 791
  (2002)

\bibitem{Robles:2012uy}
V.H. Robles, T.~Matos, Mon. Not. Roy. Astron. Soc. \textbf{422}, 282 (2012),
  \texttt{1201.3032}

\bibitem{Schive:2014hza}
H.Y. Schive, M.H. Liao, T.P. Woo, S.K. Wong, T.~Chiueh, T.~Broadhurst, W.Y.P.
  Hwang, Phys. Rev. Lett. \textbf{113}, 261302 (2014), \texttt{1407.7762}

\bibitem{2011MNRAS.413.3095H}
T.~{Harko}, \mnras \textbf{413}, 3095 (2011), \texttt{1101.3655}

\bibitem{2012MNRAS.422..282R}
V.H. {Robles}, T.~{Matos}, \mnras \textbf{422}, 282 (2012), \texttt{1201.3032}

\bibitem{2017arXiv170100912B}
T.~{Bernal}, L.M. {Fern{\'a}ndez-Hern{\'a}ndez}, T.~{Matos}, M.A.
  {Rodr{\'{\i}}guez-Meza}, ArXiv e-prints  (2017), \texttt{1701.00912}

\bibitem{0264-9381-17-1-102}
F.S. Guzman, T.~Matos, Class. Quant. Grav. \textbf{17}, L9 (2000)

\bibitem{Mbelek:2004ff}
J.P. Mbelek, Astron. Astrophys. \textbf{424}, 761 (2004),
  \texttt{gr-qc/0411104}

\bibitem{PhysRevD.69.127502}
T.H. Lee, B.J. Lee, Phys. Rev. D \textbf{69}, 127502 (2004)

\bibitem{Irsic:2017yje}
V.~Ir$\check{s}$i$\check{c}$, M.~Viel, M.G. Haehnelt, J.S. Bolton, G.D. Becker,
  Phys. Rev. Lett. \textbf{119}, 031302 (2017), \texttt{1703.04683}

\bibitem{Armengaud:2017nkf}
E.~Armengaud, N.~Palanque-Delabrouille, D.J.E. Marsh, J.~Baur, C.~Yeche, Mon.
  Not. Roy. Astron. Soc.  (2017), \texttt{1703.09126}

\bibitem{Zhang:2017chj}
J.~Zhang, J.L. Kuo, H.~Liu, Y.L.S. Tsai, K.~Cheung, M.C. Chu (2017),
  \texttt{1708.04389}

\bibitem{Mateo:1998wg}
M.~Mateo, Ann. Rev. Astron. Astrophys. \textbf{36}, 435 (1998),
  \texttt{astro-ph/9810070}

\bibitem{gilmore-2008}
G.~Gilmore, D.~Zucker, M.~Wilkinson, R.F.G. Wyse, V.~Belokurov, J.~Kleyna,
  A.~Koch, N.W. Evans, E.K. Grebel, eprint:arXiv.org:0804.1919  (2008)

\bibitem{Strigari:2008ib}
L.E. Strigari, J.S. Bullock, M.~Kaplinghat, J.D. Simon, M.~Geha, B.~Willman,
  M.G. Walker, Nature \textbf{454}, 1096 (2008), \texttt{0808.3772}

\bibitem{Daddi:2005ym}
E.~Daddi et~al., Astrophys. J. \textbf{631}, L13 (2005),
  \texttt{astro-ph/0507504}

\bibitem{2009Natur.460..717V}
P.G. {van Dokkum}, M.~{Kriek}, M.~{Franx}, \nat \textbf{460}, 717 (2009),
  \texttt{0906.2778}

\bibitem{Trujillo21112007}
I.~Trujillo, C.J. Conselice, K.~Bundy, M.C. Cooper, P.~Eisenhardt, R.S. Ellis,
  \textbf{382}, 109 (2007)

\bibitem{UrenaLopez:2002du}
L.A. Urena-Lopez, A.R. Liddle, Phys. Rev. \textbf{D66}, 083005 (2002),
  \texttt{astro-ph/0207493}

\bibitem{Lee:2015yws}
J.W. Lee, J.~Lee, H.C. Kim, arXiv:1512.02351  (2015)

\bibitem{Lee:2008mq}
J.W. Lee, S.~Lim, D.~Choi, arXiv:0805.3827  (2008)

\bibitem{CastanedaValle:2013ava}
D.~Castaneda~Valle, E.W. Mielke, Annals Phys. \textbf{336}, 245 (2013)

\bibitem{Paredes:2015wga}
A.~Paredes, H.~Michinel, Phys. Dark Univ. \textbf{12}, 50 (2016),
  \texttt{1512.05121}

\bibitem{Schwabe:2016rze}
B.~Schwabe, J.C. Niemeyer, J.F. Engels, Phys. Rev. \textbf{D94}, 043513 (2016),
  \texttt{1606.05151}

\bibitem{Guzman:2016peo}
F.S. Guzman, J.A. Gonzalez, J.P. Cruz-Perez, Phys. Rev. \textbf{D93}, 103535
  (2016), \texttt{1605.04856}

\bibitem{2009MNRAS.393.1359M}
T.~{Matos}, A.~{V{\'a}zquez-Gonz{\'a}lez}, J.~{Maga{\~n}a}, \mnras
  \textbf{393}, 1359 (2009), \texttt{0806.0683}

\bibitem{2010ApJ...721.1509R}
I.~{Rodr{\'{\i}}guez-Montoya}, J.~{Maga{\~n}a}, T.~{Matos},
  A.~{P{\'e}rez-Lorenzana}, \apj \textbf{721}, 1509 (2010), \texttt{0908.0054}

\bibitem{moffat-2006}
J.W. Moffat, astro-ph/0602607  (2006)

\bibitem{Arbey:2001qi}
A.~Arbey, J.~Lesgourgues, P.~Salati, Phys. Rev. \textbf{D64}, 123528 (2001),
  \texttt{astro-ph/0105564}

\bibitem{Matos:2009rw}
T.~Matos, arXiv:0909.3634  (2009), \texttt{0909.3634}

\bibitem{Huang:2013spa}
K.~Huang, Int. J. Mod. Phys. \textbf{A28}, 1330049 (2013), [,15(2014)],
  \texttt{1309.5707}

\bibitem{Gogberashvili:2017gru}
M.~Gogberashvili, A.S. Sakharov, arXiv  (2017), \texttt{1702.05757}

\bibitem{Lee:2015ema}
J.W. Lee, arXiv:1510.07968  (2015), \texttt{1510.07968}

\bibitem{Khmelnitsky:2013lxt}
A.~Khmelnitsky, V.~Rubakov, JCAP \textbf{1402}, 019 (2014), \texttt{1309.5888}

\bibitem{Aoki:2016mtn}
A.~Aoki, J.~Soda, Phys. Rev. \textbf{D93}, 083503 (2016), \texttt{1601.03904}

\bibitem{Aoki:2016kwl}
A.~Aoki, J.~Soda, arXiv:1608.05933  (2016)

\bibitem{Blas:2016ddr}
D.~Blas, D.L. Nacir, S.~Sibiryakov, Phys. Rev. Lett. \textbf{118}, 261102
  (2017), \texttt{1612.06789}

\bibitem{Li:2013nal}
B.~Li, T.~Rindler-Daller, P.R. Shapiro, Phys. Rev. \textbf{D89}, 083536 (2014),
  \texttt{1310.6061}

\bibitem{Kim:2015yna}
J.E. Kim, D.J.E. Marsh, Phys. Rev. \textbf{D93}, 025027 (2016),
  \texttt{1510.01701}

\end{thebibliography}


\end{document}